# Finding Hannay angle in dissipative oscillatory systems via conservative perturbation theory

Rohitashwa Chattopadhyay,[1, *] Tirth Shah,[2, 3, †] and Sagar Chakraborty[1, ‡]

[1] *Department of Physics, Indian Institute of Technology Kanpur, Uttar Pradesh 208016, India*
[2] *Max Planck Institute for the Science of Light, Staudtstraße 2, Erlangen 91058, Germany*
[3] *Department of Physics, University of Erlangen-Nürnberg, Staudtstraße 7, Erlangen 91058, Germany*

Usage of a Hamiltonian perturbation theory for a nonconservative system is counterintuitive and in general, a technical impossibility by definition. However, the time-independent dual Hamiltonian formalism for the nonconservative systems have opened the door for using various conservative perturbation theories for investigating the dynamics of such systems. Here we demonstrate that the Lie transform Hamiltonian perturbation theory can be adapted to find the perturbative solutions and the frequency corrections for the dissipative oscillatory systems. As a further application, we use the perturbation theory to analytically calculate the Hannay angle for the van der Pol oscillator's limit cycle trajectory when its parameters—the strength of the nonlinearity and the frequency of the linear part—evolve cyclically and adiabatically. For this van der Pol oscillator, we also numerically calculate the corresponding geometric phase and establish its equivalence with the Hannay angle.

## I. INTRODUCTION

In 1956 [1], Pancharatnam identified a geometrical phase while investigating the rotation of the polarization of the light beams interacting with crystals. Thereafter, there have been reports of geometric phases in various types of physical systems [2], *e.g.*, in neutron optics, nuclear magnetic resonance, quantum mechanics [3], molecular systems [4], robotics [5, 6], control theory [7], and even in stock trading [8]. The phase exclusively depends on the geometrical pathway followed by the system in the parameter space; hence the adjective 'geometrical'. Although non-adiabatic counterparts exist, the geometric phases are more traditionally recognized with the phase changes accumulated as the time-dependent, periodic parameters evolve adiabatically. The discovery of the mechanical analogue of such a phase was surprisingly delayed and is now known as the Hannay angle [9] after its discoverer.

The calculation of the Hannay angle for conservative systems is a textbook exercise [10] in which the system needs to be expressed in terms of the action-angle variables. This presents a problem for the case of dissipative systems where an action-angle formalism is not possible. However, this problem has been circumvented for a class of dissipative systems and a method to find their geometric phases has been proposed [11]. Such geometrical phases have been calculated for nonlinear dissipative systems with continuous spatial symmetries [12] and applied to laser physics [13]. This theory has been further exploited to calculate the phase shift of a wave-front in a reaction-diffusion model [14]. The geometrical phase shift has also been identified in cell division cycles [15]. Intriguingly, one can show [16] that the Hannay angle of a generalized harmonic oscillator, possessing a Hamiltonian, is related to the geometric phase of a damped harmonic oscillator with time dependent parameters. It is, however, not known if the geometric phase of any high dimensional dissipative oscillatory system is equivalent to the Hannay angle of a related conservative system. Moreover, the aforementioned method has a limited scope and cannot be extended to find the quantum mechanical analogues of the geometrical phases because of the absence of Hamiltonians.

Fortunately, Hamilton's principle of stationary action, in principle, allows for writing Lagrangians for nonconservative systems and connecting them with corresponding equations of motions via the principle of stationary action. This is so because the principle has a subtle advantage over some other action principles (*e.g.*, Maupertuis principle): it does not require energy to be conserved over the paths for which actions are calculated and extremized to arrive at the equations of motion. Only problem is while for a conservative system there is a general prescription of writing Lagrangian (*viz.*, the Lagrangian is the difference between the potential and the kinetic energies of the system), no such general prescription is available otherwise. But researchers have found some such Lagrangians and hence their Legendre duals, Hamiltonians, *e.g.*, Caldirola–Kanai Hamiltonian [17, 18] for damped simple harmonic oscillator, Hamiltonians for dissipative nonlinear oscillators via Jacobi last multiplier method [19, 20] and via extended Prelle–Singer method [21, 22], *etc.* All these are either time-dependent Hamiltonians or are time independent but globally not defined in the corresponding phase spaces. With a view to writing time independent globally defined Hamiltonians for general nonlinear systems and subsequently investigating the issue of quantization of such systems, one can improvise on the Bateman's dual Hamiltonian [23] for the damped harmonic oscillator to propose a scheme [24] for writing Hamiltonians for Liénard systems [25].

Such a scheme can be formalized by the use of Hamil-

* crohit@iitk.ac.in
† tirth.shah@fau.de
‡ sagarc@iitk.ac.in

ton's principle with initial data [26]. However, this formalism has a caveat that in order to write the Hamiltonian the phase space of the system under consideration is extended such that there are double the original number of canonically conjugate variables. As a consequence, the Hamiltonian leads to two coupled equations of motion that includes an auxiliary equation of motion describing the evolution of those variables which are not a part of the original system. In this paper, however, we put this caveat to good use by inventing an oxymoronic perturbation method. In this method, the Lie transform Hamiltonian perturbation theory (LTHPT) [27–29] has been employed to find the amplitude and the frequency perturbatively for an oscillatory state of a nonconservative Liénard system. The preceding sentence appears counterintuitive because the perturbation theory is, by definition and construction, meant to be explicitly used only for conservative Hamiltonian systems. It must be appreciated that in the context of the present paper, developing such perturbation method is of paramount practical importance because even if one manages to obtain a Hamiltonian formalism for the nonlinear limit cycle systems (and hence find the corresponding action-angle variables and the Hannay angles), one must calculate the Hannay angle via perturbation techniques because nonlinear differential equations are mostly not analytically solvable.

Another such perturbation method specific to Hamiltonian systems, *viz.*, canonical perturbation theory (see, *e.g.*, [10]), has recently been generalized [24] to accommodate the van der Pol oscillator (vdPO) [30]—which is a Liénard system that possesses a stable limit cycle solution by virtue of its nonlinear and dissipative nature—in its scope (see also [31, 32]). The main idea is to choose the initial conditions pertaining to the auxiliary equation to the vdPO — termed auxiliary van der Pol oscillator — in such a way that the inherent problem of small denominators is avoided order by order upto second order in nonlinearity. Such a liberty with the auxiliary vdPO is possible because its unidirectional coupling with the vdPO makes it dynamically inconsequential as far as the evolution of the vdPO is concerned. But there are at least two practical limitations of this generalization: i) the method becomes analytically intractable as results for higher orders in nonlinearity are sought, and ii) even if the higher order calculations are addressed by brute force, the notorious problem of small denominators makes the method useless beyond second order.

In this paper, using the paradigmatic vdPO for the sake of concreteness and without any loss of generality, we illustrate how our method—based on the LTHPT— can yield the amplitude, the frequency, and most importantly the Hannay angle for the limit cycle systems. To this end, we first succinctly discuss the classical geometric phases in Sec. II and then proceed to develop the method in Sec. III. Before concluding in Sec. V, we find the Hannay angle perturbatively in Sec. IV by employing the method on the vdPO when its parameters are time-dependent. We also establish the angle's equivalence with the corresponding geometric phase associated with the limit cycle systems.

## II. CLASSICAL GEOMETRIC PHASES

Consider a system with '$d$' degrees of freedom, described by time dependent Hamiltonian $H(\mathbf{q}, \mathbf{p}; \boldsymbol{\lambda}(t))$, where $\mathbf{q} = \{q_1, q_2, .., q_d\}$ are the generalized coordinates, $\mathbf{p} = \{p_1, p_2, .., p_d\}$ are the generalized momenta, and $\boldsymbol{\lambda}(t) = \{\lambda_1(t), \lambda_2(t), ..., \lambda_N(t)\}$ are $N \geq 2$ time dependent parameters. When the parameters are constant, the system is assumed to be completely integrable which implies that one can express the Hamiltonian in terms of the action-angle variables, $(\mathbf{I}, \boldsymbol{\phi})$. The total change over a time $T$ in one of the angle variables, when the parameters are allowed to change with time, can be written as:

$$\Delta\phi = \int_0^T \omega(\boldsymbol{\lambda}(t))\, dt + \int_0^T dt\, \dot{\boldsymbol{\lambda}} \cdot \left(\frac{\partial\phi}{\partial\boldsymbol{\lambda}}\right)_{\mathbf{q},\mathbf{p}}. \quad (1)$$

Here, $\omega$ is the corresponding frequency of the system with the parameters held constant. The first term in Eq. (1) is known as the dynamic phase. It is the second term, which is of our interest, is almost impossible to calculate except if *a priori* knowledge of $\phi$ as a function of $\boldsymbol{\lambda}$ is not known. However, if the parameters are varied adiabatically and periodically with a time period $T \gg 2\pi/\omega$, then one can approximate this term by averaging to define Hannay angle ($\phi_H$) as follows:

$$\phi_H \equiv \oint \boldsymbol{A} \cdot d\boldsymbol{\lambda}, \quad (2)$$

where,

$$\boldsymbol{A} \equiv \frac{1}{(2\pi)^d} \int_0^{2\pi} \int_0^{2\pi} ... \int_0^{2\pi} \prod_{i=1}^d d\phi_i \left(\frac{\partial\phi}{\partial\boldsymbol{\lambda}}\right)_{\mathbf{q},\mathbf{p}}. \quad (3)$$

Note that we have closed a loop in the parameter space to obtain the Hannay angle ($\phi_H$).

It is evident that the action-angle variables and hence, the Hamiltonian formulation is of the central most importance in obtaining the Hannay angle for a given system. This presents a problem for case of dissipative systems where such a Hamiltonian formulation is difficult to obtain. Nevertheless, a geometric phase for dissipative, oscillatory systems with limit cycles has been proposed in [11], and numerically obtained for various models of chemical oscillators [33, 34]. We now very briefly discuss the geometric phase for a two dimensional dissipative system (*viz.*, the vdPO) possessing limit cycles. Such a system with slowly varying parameters is represented by the following set of differential equations:

$$\frac{dr}{dt} = f(r, \theta, \boldsymbol{\lambda}), \quad (4a)$$

$$\frac{d\theta}{dt} = \Omega(r, \theta, \boldsymbol{\lambda}), \quad (4b)$$

in radial-polar coordinates.

Let the equation for the limit cycle, which exists for constant $\boldsymbol{\lambda}$, be given by $r = R(\theta, \boldsymbol{\lambda})$. The parameters are changing adiabatically and periodically with the time period $T$, i.e., $1/\omega \ll T$. Therefore, the dynamics about the limit cycle can be described by the variable $z = r - R(\theta, \boldsymbol{\lambda})$ that denotes the small deviations about the limit cycle. It can further be shown [11] that $z = \dot{\boldsymbol{\lambda}} \cdot \boldsymbol{\zeta}(\theta, \boldsymbol{\lambda})$ upto first order in $(\omega T)^{-1}$, where $\boldsymbol{\zeta}$ is a vector function with no explicit time dependence. Thus, upto the first order in $z$, Eq. (4b) becomes

$$\frac{d\theta}{dt} = \Omega(\theta, \boldsymbol{\lambda}) + z\frac{\partial \Omega}{\partial z}(\theta, \boldsymbol{\lambda}), \tag{5}$$

where $\Omega(\theta, \boldsymbol{\lambda}) \equiv \Omega(R(\theta, \boldsymbol{\lambda}), \theta, \boldsymbol{\lambda})$.

Now, it is customary to use a reparametrizing variable ($\psi$) such that the angular frequency is well-defined and constant (for a fixed value of $\boldsymbol{\lambda}$) upto leading order in $z$, i.e., we define

$$\psi(\theta, \boldsymbol{\lambda}) \equiv \omega(\boldsymbol{\lambda}) \int_0^\theta \frac{d\theta'}{\Omega(\theta', \boldsymbol{\lambda})}. \tag{6}$$

Here, $\omega(\boldsymbol{\lambda}) \equiv 2\pi[\int_0^{2\pi} d\theta/\Omega(\theta, \boldsymbol{\lambda})]^{-1}$ is the frequency of the limit cycle trajectory when the parameters are held constant. Hence,

$$\dot{\psi} = \frac{\partial \psi}{\partial \theta}\dot{\theta} + \frac{\partial \psi}{\partial \boldsymbol{\lambda}} \cdot \dot{\boldsymbol{\lambda}}, \tag{7}$$

$$\implies d\psi = \omega(\boldsymbol{\lambda})dt + d\boldsymbol{\lambda} \cdot \boldsymbol{\zeta}\omega(\boldsymbol{\lambda})\frac{\partial \ln \Omega}{\partial z}(\theta, \boldsymbol{\lambda}) + \frac{\partial \psi}{\partial \boldsymbol{\lambda}}, \tag{8}$$

where we make use of Eq. (5). The dynamic phase is given by $\int_0^T \omega(\boldsymbol{\lambda})dt$. The geometric phase ($\psi_G$) is obtained from the remaining two terms in the Eq. (8) which are averaged over a full cycle parametrized by $\psi$ followed by an integration over the closed loop in the parameter space, i.e.,

$$\psi_G = \oint \mathbf{A}^{(diss)} \cdot d\boldsymbol{\lambda}, \tag{9}$$

where,

$$\mathbf{A}^{(diss)} \equiv \frac{\omega(\boldsymbol{\lambda})}{2\pi} \int_0^{2\pi} \frac{d\theta}{\Omega(\theta, \boldsymbol{\lambda})}\left(\boldsymbol{\zeta}\omega(\boldsymbol{\lambda})\frac{\partial \ln \Omega}{\partial z}(\theta, \boldsymbol{\lambda}) + \frac{\partial \psi}{\partial \boldsymbol{\lambda}}\right). \tag{10}$$

It is intriguing to note that, while the second term in the right hand side of Eq. (10) also appears in phase shifts for one-dimensional flows on a circle, the first term is exclusive to two-dimensional limit cycle systems—which are of our main interest in this paper.

Now consider a case where we fortunately have some sort of Hamiltonian formulation for a dissipative oscillatory system. Then a situation arises where both $\psi_G$ and $\phi_H$ can exist when the parameters are varying adiabatically. Is there a relation between them? We anticipate that the answer to this question is in affirmative because for the case of one dimensional flow on a circle, it has been shown [35] that, in principle, the geometric phase follows from the canonical equations of motion and can be identified with the Hannay angle of the underlying Hamiltonian system. However, it is non-trivial to formally prove similar connection generically for higher dimensional systems. In this paper, we confine ourselves in trying to illustrate similar connection for two dimensional Liénard systems possessing limit cycles by taking the example of the vdPO. It must also be emphasized that the vdPO, along with its variations, has been very useful in modelling several realistic systems across disciplines, e.g., physics [36–39], chemistry [40], biology [41, 42], mathematics [43], economics [44], and seismology [45]. It must be emphasized that in our endeavour we have to surmount the technical obstacle of calculating the Hannay angle perturbatively because most often (e.g., for the vdPO) we cannot solve the equations exactly. To this end, in the next section, after utilizing the auxiliary vdPO to construct a Hamiltonain formalism for the vdPO, we lay out the procedure of the LTHPT tuned to our purpose.

## III. LTHPT FOR THE vdPO

The LTHPT is a specific type of Lie transform perturbation theory wherein, like in canonical perturbation theory, canonically conjugate variables of the Hamiltonian systems are explicitly put into use. Its variational formulation has also been developed [46]. It may be mentioned that the symplectic structure of Hamiltonian systems has been utilized in developing other methods, e.g., one that handle strongly perturbed systems [47]. Although debatable [48, 49], it is generally believed that Hamiltonian perturbation theories are advantageous in comparison to non-Hamiltonian perturbation theories because in contrast to finding normal forms of vector fields in the later, one only has to formally deal with a scalar function (Hamiltonian) in the former.

It must be remarked that the idea of introducing auxiliary equation is a well-established concept. In fact, it can be shown [32] that any $n$-dimensional non-conservative motion can be imbedded in a $2n$-dimensional space to arrive at a canonical structure for the extended system. Specifically, if there is an arbitrary $n$-dimensional autonomous system $\dot{\mathbf{x}} = \mathbf{f}(\mathbf{x})$ ($\mathbf{x}$ being the position vector of a phase point and $\mathbf{f}$ being a smooth phase velocity vector field), it can be augmented by introducing another coupled $n$-dimensional equation $\dot{\mathbf{y}} = -[D\mathbf{f}(\mathbf{x})]^T\mathbf{y}$ so that the entire $2n$-dimensional system admits a Hamiltonian, $H = \mathbf{y}^T\mathbf{f}(\mathbf{x})$ ($\mathbf{y}^T$, is transpose of $\mathbf{y}$). However, such an extended system and, thus, its Hamiltonian are not unique. Though this Hamiltonian when used for the perturbative analysis of the vdPO doesn't explicitly encounter the problem of small denominators, the Hamiltonian given by Eq. (13) does. The main advantage of the latter Hamiltonian is that it can be seen as an extension of Bateman Hamiltonian which forms the basis of several attempts [50, 51] of developing a formalism of quantizing dissipative systems. Hence, how to use this particular Hamiltonian for carrying out the LTHPT while confronting the problem of small denominators is not only an open problem but also of important research interest.

Without further ado, we now employ the LTHPT on the vdPO and discuss in fairly detailed fashion how the Hamiltonian perturbation theory can be employed for the dissipative system.

This mathematical process plays a pivotal role in obtaining the action angle variable of our system in the form of a series. As a side result we shall obtain the frequency correction and the amplitude correction for the vdPO upto $\mathcal{O}\left(\varepsilon^4\right)$ and $\mathcal{O}\left(\varepsilon^3\right)$ respectively. The vdPO can be mathematically described by the following second order differential equation:

$$\ddot{x}(t) + \varepsilon\left[x(t)^2 - 1\right]\dot{x}(t) + \omega^2 x(t) = 0, \quad (11)$$

where $\omega$ can be considered as the unperturbed frequency when the nonlinearity is absent i.e., $\varepsilon = 0$. It possesses a limit cycle of amplitude 2 units and frequency $\omega$ for small positive $\varepsilon$. Naturally, the presence of the attractor means that the system is dissipative and cannot be described globally by a time independent Hamiltonian. However, by extending the phase space of the vdPO by including a unidirectionally driven auxiliary system mathematically given by

$$\ddot{y}(t) - \varepsilon\left[x(t)^2 - 1\right]\dot{y}(t) + \omega^2 y(t) = 0, \quad (12)$$

a Hamiltonian formulation can be effected.

Note that Eq. (11) and Eq. (12) describe how a phase point $(x, \dot{x}, y, \dot{y})$ moves in a four-dimensional Eucledian space; the projection of the corresponding phase trajectory on $(x, \dot{x})$ plane represents the solution of the vdPO. Most importantly the auxiliary vdPO equation (Eq. (12)) is dynamically inconsequential for our purpose as it does not give any feedback to the vdPO. Both the equations for the vdPO and the auxiliary vdPO are obtainable from the following single Hamiltonian:

$$H\left(x, y, p_x, p_y\right) = p_x p_y + \omega^2 xy + \varepsilon\left(x^2 - 1\right)y p_y. \quad (13)$$

Here, the corresponding Lagrangian is $L = \dot{x}\dot{y} - \omega^2 xy - \varepsilon\left(x^2 - 1\right)\dot{x}y$ and hence the generalised momenta are given by $p_x = \partial L/\partial\dot{x}$ and $p_y = \partial L/\partial\dot{y}$.

Coming back to our main calculations, we now perform a canonical transformation $(x, y, p_x, p_y) \to (X, Y, P_X, P_Y)$ using a generating function $F_2\left(x, y, P_X, P_Y\right) = P_X(x+y)/\sqrt{2} + P_Y(x-y)/\sqrt{2}$ to write the unperturbed part of Hamiltonian in terms of a Hamiltonian resembling the one for two uncoupled harmonic oscillators. The new Hamiltonian $H\left(X, Y, P_X, P_Y\right)$ is found to be

$$H\left(X, Y, P_X, P_Y\right) = \frac{P_X^2}{2} + \omega^2\frac{x^2}{2} - \left(\frac{P_Y^2}{2} + \omega^2\frac{Y^2}{2}\right) \\ + \varepsilon\left(\frac{P_X - P_Y}{\sqrt{2}}\right)\left(\frac{X - Y}{\sqrt{2}}\right)\left[\frac{1}{2}\left(X + Y\right)^2 - 1\right]. \quad (14)$$

In the LTHPT, the Hamiltonian of Eq. (14) is initially written in terms of $\left(\alpha_1^{(0)}, \alpha_2^{(0)}, \beta_1^{(0)}, \beta_2^{(0)}\right)$ coordinates which are obtained by solving the Hamilton–Jacobi equation for the unperturbed $(\varepsilon = 0)$ Hamiltonian. The new variables $\left(\alpha_1^{(0)}, \alpha_2^{(0)}, \beta_1^{(0)}, \beta_2^{(0)}\right)$ are related to old variables $(X, Y, P_X, P_Y)$ by the following relations:

$$X = \frac{\sqrt{2\alpha_1^{(0)}}}{\omega}\sin\left[\omega\left(t + \beta_1^{(0)}\right)\right], \quad (15a)$$

$$Y = \frac{\sqrt{2\alpha_2^{(0)}}}{\omega}\sin\left[\omega\left(t - \beta_2^{(0)}\right)\right], \quad (15b)$$

$$P_X = \sqrt{2\alpha_1^{(0)}}\cos\left[\omega\left(t + \beta_1^{(0)}\right)\right], \quad (15c)$$

$$P_Y = -\sqrt{2\alpha_2^{(0)}}\cos\left[\omega\left(t - \beta_2^{(0)}\right)\right]. \quad (15d)$$

The transformed Hamiltonian $H\left(\alpha_1^{(0)}, \alpha_2^{(0)}, \beta_1^{(0)}, \beta_2^{(0)}, t\right)$ thus is expressible as a formal series,

$$H\left(\alpha_1^{(0)}, \alpha_2^{(0)}, \beta_1^{(0)}, \beta_2^{(0)}, t\right) = \sum_{n=0}^{\infty}\frac{\varepsilon^n}{n!}H_n, \quad (16)$$

where $H_i = \delta_{i1}H_1$, $\delta_{i1}$ being the standard Kronecker delta. A near-identity transformation is introduced from the variables $\left(\alpha_1^{(0)}, \alpha_2^{(0)}, \beta_1^{(0)}, \beta_2^{(0)}\right)$ to another set of variables $(\alpha_1, \alpha_2, \beta_1, \beta_2)$. Let the generating function of infinitesimal canonical transformation, $S\left(\alpha_1, \alpha_2, \beta, \beta_2, t\right)$, result in Hamiltonian $K\left(\alpha_1, \alpha_2, \beta_1, \beta_2, t\right)$. We also write

$$K\left(\alpha_1, \alpha_2, \beta_1, \beta_2, t\right) = \sum_{n=0}^{\infty}\varepsilon^n\frac{K_n}{n!}, \quad (17a)$$

$$S\left(\alpha_1, \alpha_2, \beta_1, \beta_2, t\right) = \sum_{n=0}^{\infty}\varepsilon^n\frac{S_{n+1}}{n!}. \quad (17b)$$

Upto $\mathcal{O}\left(\varepsilon^4\right)$, i.e. fourth order in $\varepsilon$, $H_n$ and $K_n$ are related as [52]:

$$K_0 = H_0 = 0, \quad (18a)$$

$$K_1 = H_1 - \frac{DS_1}{Dt}, \quad (18b)$$

$$K_2 = H_2 + \hat{L}_1 H_1 + \hat{G}_1 K_1 - \frac{DS_2}{Dt}, \quad (18c)$$

$$K_3 = H_3 + \hat{L}_1 H_2 + 2\hat{L}_2 H_1 + 2\hat{G}_1 K_2 + \hat{G}_2 K_1 \\ - \frac{DS_3}{Dt}, \quad (18d)$$

$$K_4 = H_4 + \hat{L}_1 H_3 + 3\hat{L}_2 H_2 + 3\hat{L}_3 H_1 + 3\hat{G}_1 K_3 \\ + 3\hat{G}_2 K_2 + \hat{G}_3 K_1 - \frac{DS_4}{Dt}. \quad (18e)$$

Here

$$\frac{DS_n}{Dt} = \frac{\partial S_n}{\partial t} - \hat{L}_n H_0 \, ; \quad n \geq 1, \tag{19a}$$

$$\hat{L}_n f = \sum_{i=1}^{2} \frac{\partial f}{\partial \beta_i} \frac{\partial S_n}{\partial \alpha_i} - \frac{\partial f}{\partial \alpha_i} \frac{\partial S_n}{\partial \beta_i}, \tag{19b}$$

$$\hat{G}_1 = \hat{L}_1, \tag{19c}$$

$$\hat{G}_2 = \hat{L}_2 - \hat{L}_1^2, \tag{19d}$$

$$\hat{G}_3 = \hat{L}_3 - \hat{L}_1 \left( \hat{L}_2 - \hat{L}_1^2 \right) - 2\hat{L}_2\hat{L}_1. \tag{19e}$$

In Eqs. (18b)-(18e), $K_n$ and $S_n$ are unknowns.

According to the LTHPT, $K_n$ is chosen to contain long-period terms, while $S_n$ contains short-period terms. Here, long and short periods are with respect to the period of the oscillatory dynamics described by the unperturbed Hamiltonian. A standard procedure [29] is employed to obtain $K_n$ and $S_n$, $e.g.$, in Eq. (18b) $K_1$ is obtained by averaging the equation over the time period of the unperturbed oscillatory state. Consequently, $S_1$ is obtained after solving the first order partial differential equation that results on putting the value of $K_1$ back in Eq. (18b). The final objective of the method is to make the transformed Hamiltonian integrable, meaning K should be a function of only $\alpha_1$ and $\alpha_2$. However, as illustrated in Appendix A, if any resonance condition is satisfied by the Hamiltonian ($e.g.$, two constituent frequencies of the unperturbed part are same), the problem of small denominators appear making $S_1$ (for that matter any $S_n$) divergent and aperiodic. In a bid to get rid of the divergences in $S_1$, the long period secular terms in the R.H.S. of equation $\partial S_1/\partial t = H_1 - K_1$ may be absorbed in $K_1$. Unfortunately, such a scheme yields $K_1$ (and similarly $K_n$) that is a function of $\beta_1$ and $\beta_2$ as well. Clearly, if one can propose a scheme to bypass this problem to reach at the correct perturbative results for the vdPO's limit cycle, then that would be an useful result. In what follows, we accomplish exactly this.

To this end, we try to see whether there exists an initial condition for which it is possible to solve the canonical equations of motion analytically. For convenience, the equations of motion are represented symbolically as

$$\dot{\alpha}_1 = -\frac{\partial K}{\partial \beta_1} = f_1\left(\alpha_1, \alpha_2, \beta_1, \beta_2\right), \tag{20a}$$

$$\dot{\beta}_1 = +\frac{\partial K}{\partial \alpha_1} = f_2\left(\alpha_1, \alpha_2, \beta_1, \beta_2\right), \tag{20b}$$

$$\dot{\alpha}_2 = -\frac{\partial K}{\partial \beta_2} = f_3\left(\alpha_1, \alpha_2, \beta_1, \beta_2\right), \tag{20c}$$

$$\dot{\beta}_2 = +\frac{\partial K}{\partial \alpha_2} = f_4\left(\alpha_1, \alpha_2, \beta_1, \beta_2\right). \tag{20d}$$

The exact expressions are given in Appendix B. Henceforth, in this section, all the equations are written upto $\mathcal{O}\left(\varepsilon^4\right)$ unless otherwise specified. Now the plan is to make use of the expressions of $f_1, f_2, f_3$ and $f_4$ as explained below in order to choose an initial condition for which it is convenient to solve Eqs. (20a)-(20d).

We observe that $f_1 = f_3$ which implies if the initial condition is chosen to be such that

$$\alpha_1\left(0\right) = \alpha_2\left(0\right), \tag{21}$$

then we conclude, after subtracting Eq. (20a) from Eq. (20c),

$$\alpha_1\left(t\right) = \alpha_2\left(t\right). \tag{22}$$

Using the immediately preceding relation, the four first order differential equations (20a)-(20d) reduce to the following three independent first order differential equations:

$$\dot{\alpha}_1 = g_1\left(\alpha_1, \beta_1, \beta_2\right) = f_1\left(\alpha_1, \alpha_1, \beta_1, \beta_2\right), \tag{23a}$$

$$\dot{\beta}_1 = g_2\left(\alpha_1, \beta_1, \beta_2\right) = f_2\left(\alpha_1, \alpha_1, \beta_1, \beta_2\right), \tag{23b}$$

$$\dot{\beta}_2 = g_4\left(\alpha_1, \beta_1, \beta_2\right) = f_4\left(\alpha_1, \alpha_1, \beta_1, \beta_2\right). \tag{23c}$$

Next, we note that if we choose the initial conditions such that

$$\beta_1\left(0\right) = -\beta_2\left(0\right), \tag{24}$$

then it can be shown that

$$\beta_1\left(t\right) = -\beta_2\left(t\right). \tag{25}$$

This can be proved by the change of variables $\left(\alpha_1, \beta_1, \beta_2\right) \rightarrow \left(\alpha_1, \gamma_1, \gamma_2\right)$ in Eqs. (23a)-(23c), where $\gamma_1 = \beta_1 + \beta_2$ and $\gamma_2 = \beta_1 - \beta_2$, the equations reduces to the following form

$$\dot{\gamma}_1 = h_1\left(\alpha_1, \gamma_1, \gamma_2\right) = h_{10}\left(\alpha_1, \gamma_1, \gamma_2\right)\sin\gamma_1, \tag{26a}$$

$$\dot{\gamma}_2 = h_2\left(\alpha_1, \gamma_1, \gamma_2\right), \tag{26b}$$

$$\dot{\alpha}_1 = h_3\left(\alpha_1, \gamma_1, \gamma_2\right). \tag{26c}$$

Hence, $\gamma_1\left(0\right) = \beta_1\left(0\right) + \beta_2\left(0\right) = 0$ is a fixed point of the $\gamma_1$'s flow. We remark here that $h_{10}$ is finite for $\gamma_1 = 0$.

Owing to Eqs. (22) and (25), Eqs. (20a)-(20d) reduce to two unidirectionally coupled first order differential equations which are found to be

$$\dot{\alpha}_1 = \left(\alpha_1 - \frac{\alpha_1^2}{\omega^2}\right)\varepsilon + \left(\frac{-9\alpha_1^4 + 16\alpha_1^3\omega^2 - 8\alpha_1^2\omega^4}{32\omega^8}\right)\varepsilon^3, \tag{27a}$$

$$\dot{\beta}_1 = \left(\frac{-11\alpha_1^2 + 12\alpha_1\omega^2 - 2\omega^4}{16\omega^6}\right)\varepsilon^2 + \left(-\frac{2527\alpha_1^4}{3072\omega^{12}}\right.$$
$$\left. + \frac{5032\alpha_1^3\omega^2 - 3052\alpha_1^2\omega^4 + 528\alpha_1\omega^6 - 24\omega^8}{3072\omega^{12}}\right)\varepsilon^4. \tag{27b}$$

These two equations can, in principle, be solved by quadrature. In order to appreciate the meaning of Eqs. (27a) and (27b) in the context of the dynamics of original variables $(x, \dot{x})$ of the vdPO, we first need to write relationship between the initial coordinate values (as in Eqs. (21) and (24)) in terms of original variables $(x, y, \dot{x}, \dot{y})$ of the dual vdPO system. Recall that the vdPO and the auxiliary vdPO are unidirectionally coupled, i.e., the dynamics of the vdPO is not affected by the dynamics of the auxiliary vdPO. Hence, no initial condition for the auxiliary vdPO alters the dynamics of the isolated vdPO.

It is easily shown that the specific choice of initial conditions — $y(0) = 0$ and $\dot{y}(0) = 0$ — for the auxiliary vdPO corresponds to

$$X(0) = Y(0), P_X(0) = -P_Y(0) , \qquad (28)$$

which means for $\left(\alpha_1^{(0)}, \alpha_2^{(0)}, \beta_1^{(0)}, \beta_2^{(0)}\right)$:

$$\alpha_1^{(0)}(0) = \alpha_2^{(0)}(0) , \quad \beta_1^{(0)}(0) = -\beta_2^{(0)}(0) . \qquad (29)$$

Now, by writing the variables $(\alpha_1, \alpha_2, \beta_1, \beta_2)$ in terms of the variables $\left(\alpha_1^{(0)}, \alpha_2^{(0)}, \beta_1^{(0)}, \beta_2^{(0)}\right)$ using the inverse transformation algorithm of Lie transforms, one concludes that Eqs. (21) and (24) are satisfied when $y(0) = \dot{y}(0) = 0$. Summarizing, we have essentially exercised our freedom in choosing an arbitrary initial condition for the auxiliary vdPO, and shown that the selection of its fixed point as the preferred initial condition leads to Eqs. (27a) and (27b).

The fixed point of Eq. (27a) is

$$\alpha_1 = \alpha_1(t) = \omega^2 - \frac{\varepsilon^2}{32} + \mathcal{O}\left(\varepsilon^4\right) , \qquad (30)$$

apart from the trivial one, $\alpha_1 = 0$. The physical meaning of the fixed points is easily understandable in the phase space of the vdPO and the auxiliary vdPO. Note that we have already chosen the full system to be at the fixed point of the auxiliary vdPO. Therefore we need to concern ourselves only with the phase space of the vdPO. Fixed point $\alpha_1 = 0$ corresponds to the fixed point (focus/spiral) at the origin of phase space for the vdPO. The more interesting fixed point is the one given by Eq. (30). This must correspond to the limit cycle in the vdPO's phase space. Recall that we are interested in the corrections to the frequency of this limit cycle due to nonlinearity. Coming back to our calculations, substituting Eq. (30) in Eq. (27b) leads to

$$\dot{\beta}_1 = -\frac{\varepsilon^2}{16\omega^2} + \frac{17\varepsilon^4}{3072\omega^4} . \qquad (31)$$

Now, using Eqs. (15a)-(15d) and the canonical relation $(x, y, p_x, p_y) \to (X, Y, P_X, P_Y)$, we obtain

$$x = \frac{\sqrt{\alpha_1^{(0)}}}{\omega} \sin\left[\omega\left(t + \beta_1^{(0)}\right)\right] + \frac{\sqrt{\alpha_2^{(0)}}}{\omega} \sin\left[\omega\left(t - \beta_2^{(0)}\right)\right] . \qquad (32)$$

Using Eq. (22), Eq. (25) and Eq. (30), the analytical solution for the vdPO upto $\mathcal{O}\left(\varepsilon^3\right)$ is obtained by rewriting Eq. (32) in terms of variables $(\alpha_1, \alpha_2, \beta_1, \beta_2)$:

$$\begin{aligned} x(t) = {}& 2\sin B_1 - \varepsilon\left(\frac{\cos 3B_1}{4\omega}\right) \\ & + \varepsilon^2\left(\frac{\sin B_1}{64\omega^2} + \frac{3\sin 3B_1}{32\omega^2} - \frac{5\sin 5B_1}{96\omega^2}\right) \\ & + \varepsilon^3\left(\frac{13\cos B_1}{256\omega^3} + \frac{15\cos 3B_1}{512\omega^3} - \frac{85\cos 5B_1}{2304\omega^3}\right. \\ & \left. + \frac{7\cos 7B_1}{576\omega^3}\right) . \end{aligned} \qquad (33)$$

Here, $B_1 = \omega(t + \beta_1)$. From the immediately preceding equation, the frequency of the limit cycle of the vdPO comes out to be

$$\dot{B}_1 = \omega\left(1 + \dot{\beta}_1\right) = \omega - \frac{\varepsilon^2}{16\omega} + \frac{17\varepsilon^4}{3072\omega^3} + \mathcal{O}(\varepsilon^5) . \qquad (34)$$

These frequency corrections would be exactly in conformity with the results obtained using other perturbation methods [29] like Poincaré-Lindstedt technique, multiple timescales method, equivalent linearization [53], *etc.*

## IV. HANNAY ANGLE VIA THE LTHPT

We are now prepared to apply the LTHPT to calculate the Hannay angle for the limit cycle trajectory of the vdPO, using a time independent Hamiltonian given by Eq (14). We, however, must use the action-angle variables instead of the canonical coordinates used in the previous section because the Hannay angle is defined through them.

We allow $\boldsymbol{\lambda} = \{\lambda_1, \lambda_2\} \equiv \{\omega, \varepsilon\}$ to be the time dependent parameters. The vdPO with time dependent $\varepsilon$ is interesting in its own right [54]. For the purpose in hand, we assume $\omega$ and $\varepsilon$ to adiabatically change with time. Once a Hamiltonian formalism for the vdPO is obtained, we can calculate the Hannay angle via the standard technique mentioned in Sec. II. Therefore, our next task is to calculate the action-angle variables for the dual vdPO system when the parameters are constant. The primary challenge towards this objective is that the vdPO system cannot be solved analytically and therefore, the angle variables for the entire system are attained perturbatively as a power series in $\varepsilon$. The terms of this series will be calculated upto $\mathcal{O}\left(\varepsilon^2\right)$ via the LTHPT [29].

We proceed to write the Hamiltonian of Eq. (14) in terms of the action-angle variables $\left(I_1^{(0)}, I_2^{(0)}, \phi_1^{(0)}, \phi_2^{(0)}\right)$ of the unperturbed Hamiltonian. As outlined in the preceding section, the action-angle variables for the entire system, $(I_1, I_2, \phi_1, \phi_2)$, are related to the variables $\left(I_1^{(0)}, I_2^{(0)}, \phi_1^{(0)}, \phi_2^{(0)}\right)$ via a near-identity transformation and we again assume that the new Hamiltonian and the generating function can be expressed as a series [as in Eqs. (17a) and (17b)].

The canonical equations of motion, i.e., equations analogous to Eqs. (20a)-(20d), can be obtained by following the procedure outlined in the preceding section. As stated earlier, these equations are solvable and the auxiliary system is decoupled when the initial conditions are chosen such that: $y(0) = 0$ and $\dot{y}(0) = 0$, which means for the variables $\left(I_1^{(0)}, I_2^{(0)}, \phi_1^{(0)}, \phi_2^{(0)}\right)$:

$$\phi_1^{(0)}(0) + \phi_2^{(0)}(0) = 0, \sqrt{I_1^{(0)}(0)} + \sqrt{I_2^{(0)}(0)} = 0. \quad (35)$$

The above conditions imply

$$\phi_1^{(0)}(t) + \phi_2^{(0)}(t) = 0, \sqrt{I_1^{(0)}(t)} + \sqrt{I_2^{(0)}(t)} = 0, \quad (36)$$

$$\phi_1(t) + \phi_2(t) = 0, \sqrt{I_1} + \sqrt{I_2} = 0. \quad (37)$$

upto $\mathcal{O}\left(\varepsilon^4\right)$. After considering the above initial conditions, the canonical equations of motion upto $\mathcal{O}\left(\varepsilon^3\right)$ can be obtained as

$$\dot{I}_1 = \left(I_1 - \frac{I_1^2}{\omega}\right)\varepsilon + \left(\frac{-9I_1^4 + 16I_1^3\omega - 8I_1^2\omega^2}{32\omega^5}\right)\varepsilon^3, \quad (38)$$

$$\dot{\phi}_1 = \omega - \left(\frac{11I_1^2 - 12I_1\omega + 2\omega^2}{16\omega^3}\right)\varepsilon^2. \quad (39)$$

The non-zero real fixed point of Eq. (38) is

$$I_1 = I_1(t) = \omega - \frac{\varepsilon^2}{32\omega} + \mathcal{O}\left(\varepsilon^3\right). \quad (40)$$

It is instructive to note that Eqs. (38), (39), and (40) are similar to Eqs. (27a), (27b), and (30) respectively.

Before embarking on the calculation of the Hannay angle for the limit cycle trajectory of the vdPO, we remind ourselves of the set of variable transformation performed till now, i.e.,

$$(x, y, p_x, p_y) \rightarrow (X, Y, P_X, P_Y) \rightarrow \left(\phi_1^{(0)}, \phi_2^{(0)}, I_1^{(0)}, I_2^{(0)}\right)$$
$$\rightarrow (\phi_1, \phi_2, I_1, I_2). \quad (41)$$

The transformation of variables $(x, y, p_x, p_y) \leftrightarrow \left(\phi_1^{(0)}, \phi_2^{(0)}, I_1^{(0)}, I_2^{(0)}\right)$ requires simple substitution of the variables corresponding to the relations between adjacent variables set. The transformation $\left(\phi_1^{(0)}, \phi_2^{(0)}, I_1^{(0)}, I_2^{(0)}\right) \leftrightarrow (\phi_1, \phi_2, I_1, I_2)$ are obtained perturbatively as a power series in $\varepsilon$ via the algorithm of Lie transforms [29].

Now, we follow the procedure outlined in Sec. II for the calculation of the Hannay angle: First, the angle variable $\phi_1$ are written in terms of the action-angle variables of the unperturbed Hamiltonian, i.e., $(I_1^{(0)}, I_2^{(0)}, \phi_1^{(0)}, \phi_2^{(0)})$. A further change of variables from $(I_1^{(0)}, I_2^{(0)}, \phi_1^{(0)}, \phi_2^{(0)})$ to $(x, y, p_x, p_y)$ is done before the derivatives $(\partial\phi_1/\partial\varepsilon)_{x,y,p_x,p_y}$ and $(\partial\phi_1/\partial\omega)_{x,y,p_x,p_y}$ can be calculated. Writing these derivatives in terms of $(\phi_1, \phi_2, I_1, I_2)$ via the Lie transform algorithm, the average over $\phi_1$ and $\phi_2$ is performed. Finally, the conditions given in Eqs. (37) and (40) are used to obtain $\mathbf{A} = \{A_1, A_2\}$ upto $\mathcal{O}(\varepsilon)$:

$$A_1(\omega, \varepsilon) = -\frac{\varepsilon}{8\omega^2}, \quad A_2(\omega, \varepsilon) = 0. \quad (42)$$

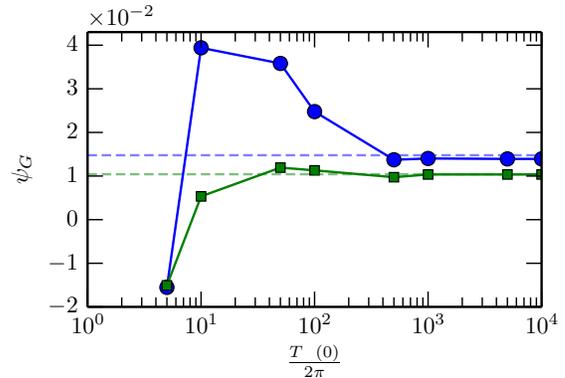

FIG. 1. *(Color online)* Equivalence of the Hannay angle ($\phi_H$) with the geometric phase ($\psi_G$) for the vdPO. $\psi_G$ is plotted against the ratio of the time period ($T$) of the parameters to the time period ($2\pi/\omega(0)$) of the system when the parameters are held fixed. The green colour and the square markers correspond to the square loop in the parameter space (see text); and the blue colour and the circular markers correspond to the elliptical loop (see text). Note that as $T\omega(0)/2\pi \rightarrow \infty$, the solid lines (aid to the eyes) for the numerical values of $\psi_G$ converge to 0.0103 and 0.013 respectively for the square and the elliptical loops. The corresponding dashed lines represents the analytically calculated $\phi_H$—0.0104 and 0.0147 respectively.

As concrete examples, we calculate the Hannay angles for the following two cases:

1. Square loop: We take a counter-clockwise square loop in parameter plane $(\omega, \varepsilon)$ such that $\omega \in (0.6, 0.8)$ and $\varepsilon \in (0.1, 0.3)$. The Hannay angle is obtained using Eq. (2) and it comes out to be 0.0104.

2. Elliptical loop: We similarly calculate the Hannay angle for an anti-clockwise elliptical loop in the parameter space where $\omega = 0.8 + 0.2\cos(2\pi t/T)$, $\epsilon = 0.1 + 0.1\sin(2\pi t/T)$, and $t \in [0, T]$. The Hannay angle comes out to be 0.0147.

Additionally, we can calculate the geometric phase, $\psi_G$ [Eq. (9)], without tapping into the Hamiltonian formalism. For that we start with Eq. (11) and numerically integrate it while simultaneously varying the parameters. Of course, it is assumed that the timescale of relaxation to the limit cycle as the parameters are varied is negligible compared to the timescale of variation of the parameters. To remove the transients we let the simulation run until the system settles (within the desired level of precision) into the limit cycle before the parameters are varied. As the time period $T$ of the parameters tend to infinity, the numerical values of the geometric phases for both the loops tend to match with the corresponding Hannay angles. Fig. 1 depicts this equivalence of the Hannay angle and the geometric phase. Therefore, it gives weight to the assertion that the geometric phase of a two-dimensional limit cycle system can be seen as a Hannay angle for the corresponding extended system possessing a Hamiltonian. More importantly, what we have achieved is that we can adapt a conservative perturbation method, the LTHPT to be precise, for analytically calculating the Hannay angle, and hence the geometric phase of a nonconservative system.

## V. DISCUSSION AND CONCLUSIONS

A time-independent Hamiltonian formalism for the damped oscillator has been formulated by Bateman [23]. In it the phase space of the system is extended by adding an auxilliary equation. The same idea of extending the phase space has been utilized few years back to propose a Hamiltonian for the vdPO unidirectionally coupled to an auxiliary equation [24]. Once the Hamiltonian is formulated, one could apply the canonical perturbation theory to calculate the frequency corrections upto second order. In this paper, we have used the same Hamiltonian but a fundamentally different Hamiltonian perturbation method, viz., the LTHPT. We have shown that it also yields similar frequency corrections for the vdPO but now we can go to even higher orders of correction analytically. While the explicit expression, Eq. (34), is well known, the formulation of the analytical technique to arrive at the results is an achievement of this paper. It is worth mentioning that since we are not interested in the dynamics of the auxiliary system, its being unidirectionally coupled to the original system allows us to choose any arbitrary initial conditions for it. This arbitrariness in the choice of the initial conditions has the benefit that it bypasses the (in)famous problem of small denominators which shows up implicitly as our inability to solve Eqs. (20a)-(20d), and ultimately yields the correct perturbative expansions for the solutions and the frequencies for the limit cycle oscillations. The other main result of this paper is that, by using LTHPT, we have successfully calculated the Hannay angle for the vdPO having adiabatically and periodically evolving time-dependent parameters. Further, we have established the equivalence between the Hannay angle and the numerically found geometric phase for the limit cycle trajectory of the vdPO. This is the first such demonstration of equivalence in a two-dimensional system.

Our adaptation of the LTHPT for nonconservative systems and its application in finding Hannay angle are useful, in principle, for any weakly nonlinear system that is described by a dual Hamiltonian à la Bateman formalism [23]. Needless to say, it means that the technique is generalizable to bring several other Liénard systems under its scope. The technique of LTHPT is a well established one that has been successfully applied to various Hamiltonian systems with weak nonlinearity. Our addition to this technique is the judicious use of the initial conditions on the auxiliary equation which helps in further simplification of this process. We again bring it to the readers' notice that it is generally suggested that, compared to the canonical perturbation theory, Lie transform perturbation theory is better suited [55] to calculate higher order corrections to unperturbed frequency; and the LTHPT has a numerical edge since it is much easily numerically implementable using a symbolic algebra package [56]. Thus, the method discussed in this paper, when applied to other relevant systems, may be more involved but one still expects to obtain fruitful results.

The method discussed in this paper can have applications in the realistic situations wherever a geometric phase is realized in a limit cycle system. For example, consider the intriguing topic of the birdsongs [57]. Here, the vocalizations may be modeled by a slightly modified vdPO [58] that describes the departure of the midpoint of the labia from the prephonatory position in a Chingolo sparrow (say), thus, modeling the generation of the various syllables of the song of the sparrow. Interestingly, two of the system-parameters—restitution constant of the labia and a linear function of the net driving pressure of the labia—are time-dependent and form a closed elliptical loop of time period 100 ms (approximately) in the parameter space. This time period is about $5 \times 10^2$ times more than the time of the oscillations of the labia. Now, although Fig. 1 has been plotted for the vdPO oscillator, we observe that an elliptical loop in its parameter space yields an almost saturated value of the geometric phase when $T\omega(0)/2\pi = 5 \times 10^2$, thus encouraging one to take up analytical calculation of the geometric phase in the birdsong. Another very interesting occurrence of the geometric phase, and hence an avenue for applying our method, is in cell cycles [15]. A Hamiltonian, its associated action angle variables, and the expression for the Hannay angle has been found for the linearized version [15] of the Goodwin's [59] of the cell cycle. Further a geometric phase has also been found for the full nonlinear Goodwin's model [60]. The geometric phase is supposed to account for the high cell-to-cell variability observed in experiments with artificial gene circuits introduced into cells. Also, owing to the finite geometric phase, a population of cells with synchronous biological oscillations to start with can become desynchronized after a few cell divisions. As yet another example we may cite the replicator-mutator dynamics of the well-known rock-paper-scissors game that possesses limit cycle oscillations [61]. Here, making the the payoff matrix elements and the rate of mutation periodic and slowly varying with time may give rise of geometric phase whose analytical evaluation may use our ideas developed in this paper. Last but not the least, the Bonhoeffer–van der Pol model—a modified version of the vdPO—for nerve membrane [62] may also make use of the methods described in our paper. This model consists of variables like excitability and refractoriness of the neuron membrane. It is quite possible that some biological process renders these parameters time dependent in such a way that the existence of the geometric phase is facilitated in the dynamics the neuron membrane.

Before ending, we remark that it is definitely of interest to find out what the quantization of the Liénard systems using the corresponding dual Hamiltonian implies. Subsequently, it would be an intriguing problem to investigate if a Berry phase [3] can be found in dissipative systems upon consistent quantization and how it would be related to the corresponding Hannay angle discussed in this paper.

## ACKNOWLEDGEMENTS

The authors thank Saheli Mitra for helpful discussions and also thank the two anonymous referees for constructive the comments that substantially improved the presentation of our paper. S.C. gratefully acknowledges financial support from the INSPIRE faculty fellowship (DST/INSPIRE/04/2013/000365) awarded by the INSA, India and DST, India.

## Appendix A: LTHPT for two nonlinearly coupled oscillators

Consider the Hamiltonian

$$H(q_1, q_2, p_1, p_2) = \frac{p_1^2}{2} + \frac{1}{2}\omega_1 q_1^2 + \frac{p_2^2}{2} + \frac{1}{2}\omega_2 q_2^2 + \varepsilon \omega_1^2 \omega_2^2 q_1^2 q_2^2. \tag{A1}$$

The above Hamiltonian describes the dynamics of two harmonic oscillators coupled by a nonlinear interaction term $\varepsilon \omega_1^2 \omega_2^2 q_1^2 q_2^2$. The nonlinearity parameter $\varepsilon$ is small and hence the dynamics can be understood by employing a perturbation theory. The Hamiltonian is initially written in terms of variables $\left(\alpha_1^{(0)}, \alpha_2^{(0)}, \beta_1^{(0)}, \beta_2^{(0)}\right)$. A near-identity transformation is performed from variables $\left(\alpha_1^{(0)}, \alpha_2^{(0)}, \beta_1^{(0)}, \beta_2^{(0)}\right)$ to variables $(\alpha_1, \alpha_2, \beta_1, \beta_2)$. Using Eq. (18b), one finds that transformed Hamiltonian $K(\alpha_1, \alpha_2, \beta_1, \beta_2, t)$ and generating function $S(\alpha_1, \alpha_2, \beta_1, \beta_2, t)$ of the infinitesimal

canonical transformation are related at the first order as:

$$K_1 = -\frac{\partial S_1}{\partial t} + \alpha_1 \alpha_2 - \alpha_1 \alpha_2 \cos\left[2\left(t + \beta_1\right)\omega_1\right]$$
$$- \alpha_1 \alpha_2 \cos\left[2\left(t + \beta_2\right)\omega_2\right]$$
$$+ \frac{\alpha_1 \alpha_2}{2} \cos\left[2\left(\omega_1 - \omega_2\right)t + 2\left(\beta_1\omega_1 - \beta_2\omega_2\right)\right]$$
$$+ \frac{\alpha_1 \alpha_2}{2} \cos\left[2\left(\omega_1 + \omega_2\right)t + 2\left(\beta_1\omega_1 + \beta_2\omega_2\right)\right]. \tag{A2}$$

$K_1$ and $S_1$ are obtained in a similar manner as that for the case of the vdPO, i.e., by employing the technique of averaging over the unperturbed system's time period. For $\omega_1$ not close to $\omega_2$, $K_1 = \alpha_1 \alpha_2$ and one can easily calculate the frequency correction by solving the canonical equations of motion. Note this is possible because there is no $\beta_{1,2}$ in $K_1$. However, if $\omega_1 \approx \omega_2$, the $\cos\left[2\left(\omega_1 - \omega_2\right)t\right]$ containing term is relatively constant over unperturbed period. Therefore, it is a long-period term and should be included in $K_1$. As a result, for $\omega_1 \approx \omega_2$, it make sense to take

$$K_1 = \alpha_1 \alpha_2 + \frac{\alpha_1 \alpha_2}{2} \cos\left[2\left(\omega_1 - \omega_2\right)t + 2\left(\beta_1\omega_1 - \beta_2\omega_2\right)\right]. \tag{A3}$$

Consequently, the canonical equations of motion upto $\mathcal{O}\left(\varepsilon\right)$ are

$$\dot{\alpha}_1 = \varepsilon \alpha_1 \alpha_2 \omega_1 \sin\left[2\left(\omega_1 - \omega_2\right)t + 2\left(\beta_1\omega_1 - \beta_2\omega_2\right)\right], \tag{A4a}$$

$$\dot{\alpha}_2 = -\varepsilon \alpha_1 \alpha_2 \omega_2 \sin\left[2\left(\omega_1 - \omega_2\right)t + 2\left(\beta_1\omega_1 - \beta_2\omega_2\right)\right], \tag{A4b}$$

$$\dot{\beta}_1 = \varepsilon \alpha_2 + \frac{\varepsilon}{2}\alpha_2 \cos\left[2\left(\omega_1 - \omega_2\right)t + 2\left(\beta_1\omega_1 - \beta_2\omega_2\right)\right], \tag{A4c}$$

$$\dot{\beta}_2 = \varepsilon \alpha_1 + \frac{\varepsilon}{2}\alpha_1 \cos\left[2\left(\omega_1 - \omega_2\right)t + 2\left(\beta_1\omega_1 - \beta_2\omega_2\right)\right]. \tag{A4d}$$

We observe that the new Hamiltonian for the case of $\omega_1 \approx \omega_2$ is no longer only a function of $\alpha_1$ and $\alpha_2$. Eqs. (A4a)-(A4d) are nonlinearly coupled and analytically intractable to solve even for the case of $\omega_1 = \omega_2$. This should be seen as the manifestation of the problem of small denominators. Alternatively but equivalently, while performing the LTHPT using action-angle variables [10], the problem of small denominators in $S_1$ is similarly tried [28] to be avoided by including in $K_1$ the terms responsible for small denominators. For example, in the case of the coupled oscillators being studied, such a term is $\cos\left[2\left(\omega_1 - \omega_2\right)t\right]$.

However, owing to the inability to solve Eqs. (A4a)-(A4d) analytically, the problem still lingers hinting that the aforementioned trick of redefining $K_1$ is not of much practical use. We remark here that for the case of the vdPO, the liberty in the choice of initial conditions of the auxiliary vdPO equation helps in successfully overcoming this very type of problem. However, for the case of the coupled oscillator we cannot use our method developed for the vdPO because none of the equations of motion, Eqs. (A4a)-(A4d), can be considered as a dynamically inconsequential auxiliary equation.

**Appendix B: Explicit expressions for the canonical equations of motion of the vdPO**

Below we provide full explicit expressions for Eq. (20a)-Eq. (20d) as obtained using MATHEMATICA software [63].

$$
\begin{aligned}
\dot{\alpha}_1 = f_1\left(\alpha_1, \alpha_2, \beta_1, \beta_2\right) = &\; \varepsilon \left[ -\frac{1}{4\omega^2}\sqrt{\alpha_1\alpha_2}\left\{ \left(\alpha_1 + \alpha_2 - 4\omega^2\right)\cos\left(\left(\beta_1 + \beta_2\right)\omega\right) + 2\sqrt{\alpha_1}\sqrt{\alpha_2}\cos\left(2\left(\beta_1 + \beta_2\right)\omega\right)\right\}\right] \\
&+ \varepsilon^2 \left[ -\frac{1}{64\omega^5}\sqrt{\alpha_1\alpha_2}\left(\alpha_1 - \alpha_2\right)\left\{22\sqrt{\alpha_1}\sqrt{\alpha_2}\cos\left(\left(\beta_1 + \beta_2\right)\omega\right) + 11\alpha_1 + 11\alpha_2 - 24\omega^2\right\}\sin\left(\left(\beta_1 + \beta_2\right)\omega\right)\right] \\
&+ \varepsilon^3 \left[ -\frac{1}{2048\omega^8}\sqrt{\alpha_1\alpha_2}\left\{ \left(9\alpha_1^3 + 54\alpha_1^2\alpha_2 - 64\omega^2\alpha_1^2 + 54\alpha_1\alpha_2^2 - 192\alpha_1\alpha_2\omega^2 + 128\alpha_1\omega^4 + 9\alpha_2^3 - 64\alpha_2^2\omega^2 + 128\alpha_2\omega^4\right)\right. \right. \\
&\quad \times \cos\left(\left(\beta_1 + \beta_2\right)\omega\right) + 2\sqrt{\alpha_1\alpha_2}\left(27\alpha_1^2 + 27\alpha_2^2 - 128\alpha_2\omega^2 + 128\omega^4 + 72\alpha_1\alpha_2 - 128\alpha_1\omega^2\right)\cos\left(2\left(\beta_1 + \beta_2\right)\omega\right) \\
&\quad \left. + 3\sqrt{\alpha_1\alpha_2}\left(27\alpha_1 + 27\alpha_2 - 64\omega^2\right)\cos\left(3\left(\beta_1 + \beta_2\right)\omega\right) + 36\sqrt{\alpha_1\alpha_2}\cos\left(4\left(\beta_1 + \beta_2\right)\omega\right)\right\}\right] \\
&+ \varepsilon^4 \left[ -\frac{1}{98304\omega^{11}}\sqrt{\alpha_1\alpha_2}\left(\alpha_1 - \alpha_2\right)\left\{2527\alpha_1^3 + 22743\alpha_1^2\alpha_2 + 22743\alpha_1\alpha_2^2 + 2527\alpha_2^3 - 15096\alpha_1^2\omega^2 - 60384\alpha_1\alpha_2\omega^2 \right. \right. \\
&\quad - 15096\alpha_2^2\omega^2 + 24416\alpha_1\omega^4 + 24416\alpha_2\omega^4 - 8448\omega^6 + 2\sqrt{\alpha_1\alpha_2}\left(7581\alpha_1^2 + 22743\alpha_1\alpha_2 + 7581\alpha_2^2 - 30192\alpha_1\omega^2 - 30192\alpha_2\omega^2\right. \\
&\quad \left. + 24416\omega^4\right)\times\cos\left(\left(\beta_1 + \beta_2\right)\omega\right) + 6\alpha_1\alpha_2\left(2527\alpha_1 + 2527\alpha_2 - 5032\omega^2\right)\cos\left(2\left(\beta_1 + \beta_2\right)\omega\right) \\
&\quad \left. \left. + 5054\alpha_1\alpha_2\sqrt{\alpha_1\alpha_2}\cos\left(3\left(\beta_1 + \beta_2\right)\omega\right)\right\}\right].
\end{aligned}
$$

$$(B1)$$

$$
\begin{aligned}
\dot{\beta}_1 = f_2\left(\alpha_1, \alpha_2, \beta_1, \beta_2\right) = &\; \varepsilon \left[ \frac{1}{8\sqrt{\alpha_1}\omega^3}\sqrt{\alpha_2}\left\{3\alpha_1 + \alpha_2 - 4\omega^2 + 4\sqrt{\alpha_1\alpha_2}\cos\left(\left(\beta_1 + \beta_2\right)\omega\right)\right\}\sin\left(\left(\beta_1 + \beta_2\right)\omega\right)\right] \\
&+ \varepsilon^2 \left[ \frac{1}{256\sqrt{\alpha_1}\omega^6}\left\{2\sqrt{\alpha_2}\left(-55\alpha_1^2 + 72\alpha_1\omega^2 + 11\alpha_2^2 - 24\alpha_2\omega^2\right)\cos\left(\left(\beta_1 + \beta_2\right)\omega\right)\right. \right. \\
&\quad \left. \left. + \sqrt{\alpha_1}\left(-33\alpha_1^2 - 66\alpha_1\alpha_2 + 33\alpha_2^2 + 96\alpha_1\omega^2 - 32\omega^4 - 22\alpha_2\left(2\alpha_1 - \alpha_2\right)\cos\left(2\left(\beta_1 + \beta_2\right)\omega\right)\right)\right\}\right] \\
&+ \varepsilon^3 \left[ \frac{1}{4096\sqrt{\alpha_1}\omega^9}\sqrt{\alpha_2}\left\{63\alpha_1^3 + 405\alpha_1^2\alpha_2 + 243\alpha_1\alpha_2^2 + 9\alpha_2^3 - 320\alpha_1^2\omega^2 - 768\alpha_1\alpha_2\omega^2 - 64\alpha_2^2\omega^2 + 384\alpha_1\omega^4 + 128\alpha_2\omega^4 \right. \right. \\
&\quad + 4\sqrt{\alpha_1\alpha_2}\left(81\alpha_1^2 + 162\alpha_1\alpha_2 + 27\alpha_2^2 - 256\alpha_1\omega^2 - 128\alpha_2\omega^2 + 128\omega^4\right)\cos\left(\left(\beta_1 + \beta_2\right)\omega\right) + 6\alpha_1\alpha_2\left(45\alpha_1 + 27\alpha_2 - 64\omega^2\right) \\
&\quad \left. \times \cos\left(2\left(\beta_1 + \beta_2\right)\omega\right) + 72\alpha_1\alpha_2\sqrt{\alpha_1\alpha_2}\cos\left(3\left(\beta_1 + \beta_2\right)\omega\right)\right\}\sin\left(\left(\beta_1 + \beta_2\right)\omega\right)\right] \\
&+ \varepsilon^4 \left[ \frac{1}{786432\sqrt{\alpha_1}\omega^{12}}\left\{4\sqrt{\alpha_2}\left(-22743\alpha_1^4 + \alpha_1^3\left(-88445\alpha_2 + 105672\omega^2\right) + 80\alpha_1^2\left(1887\alpha_2\omega^2 - 1526\omega^4\right) + \alpha_2\left(2527\alpha_2^3\right.\right.\right. \right. \\
&\quad \left. \left. - 15096\alpha_2^2\omega^2 + 24416\alpha_2\omega^4 - 8448\omega^6\right) + 3\alpha_1\left(12635\alpha_2^3 - 30192\alpha_2^2\omega^2 + 8448\omega^6\right)\right)\cos\left(\left(\beta_1 + \beta_2\right)\omega\right) \\
&\quad + \sqrt{\alpha_1}\left(-12635\alpha_1^4 - 151620\alpha_1^3\alpha_2 - 151620\alpha_1^2\alpha_2^2 + 101080\alpha_1\alpha_2^3 + 37905\alpha_2^4 + 80512\alpha_1^3\omega^2 + 483072\alpha_1^2\alpha_2\omega^2\right. \\
&\quad - 161024\alpha_2^3\omega^2 - 146496\alpha_1^2\omega^4 - 292992\alpha_1\alpha_2\omega^4 + 146496\alpha_2^2\omega^4 + 67584\alpha_1\omega^6 - 6144\omega^8 + 4\alpha_2\left(-30324\alpha_1^3\right. \\
&\quad + \alpha_1^2\left(-37905\alpha_2 + 90576\omega^2\right) + 14\alpha_1\left(1805\alpha_2^2 - 3488\omega^4\right) + \alpha_2\left(7581\alpha_2^2 - 30192\alpha_2\omega^2 + 24416\omega^4\right)\right)\cos\left(2\left(\beta_1 + \beta_2\right)\omega\right) \\
&\quad - 4\alpha_2\sqrt{\alpha_1\alpha_2}\left(17689\alpha_1^2 - 7581\alpha_2^2 - 25160\alpha_1\omega^2 + 15096\alpha_2\omega^2\right)\cos\left(3\left(\beta_1 + \beta_2\right)\omega\right) + \left(-15162\alpha_1^2\alpha_2^2 + 10108\alpha_1\alpha_2^3\right) \\
&\quad \left. \left. \times \cos\left(4\left(\beta_1 + \beta_2\right)\omega\right)\right)\right\}\right].
\end{aligned}
$$

$$(B2)$$

$$
\dot{\alpha}_2 = f_3\left(\alpha_1, \alpha_2, \beta_1, \beta_2\right) = f_1\left(\alpha_1, \alpha_2, \beta_1, \beta_2\right).
$$

$$(B3)$$

$$\dot{\beta}_2 = f_4\left(\alpha_1, \alpha_2, \beta_1, \beta_2\right) = \varepsilon\left[\frac{1}{8\sqrt{\alpha_2}\omega^3}\sqrt{\alpha_1}\left\{3\alpha_2 + \alpha_1 - 4\omega^2 + 4\sqrt{\alpha_1\alpha_2}\cos\left(\left(\beta_1 + \beta_2\right)\omega\right)\right\}\sin\left(\left(\beta_1 + \beta_2\right)\omega\right)\right]$$

$$+ \varepsilon^2\left[\frac{1}{256\sqrt{\alpha_2}\omega^6}\left\{-2\sqrt{\alpha_1}\left(-55\alpha_2^2 + 72\alpha_2\omega^2 + 11\alpha_1^2 - 24\alpha_1\omega^2\right)\cos\left(\left(\beta_1 + \beta_2\right)\omega\right)\right.\right.$$

$$\left.\left. + \sqrt{\alpha_2}\left(-33\alpha_1^2 + 66\alpha_1\alpha_2 + 33\alpha_2^2 - 96\alpha_2\omega^2 + 32\omega^4 - 22\alpha_1\left(\alpha_1 - 2\alpha_2\right)\cos\left(2\left(\beta_1 + \beta_2\right)\omega\right)\right)\right\}\right]$$

$$+ \varepsilon^3\left[\frac{1}{4096\sqrt{\alpha_2}\omega^9}\sqrt{\alpha_1}\left\{63\alpha_2^3 + 405\alpha_1\alpha_2^2 + 243\alpha_1^2\alpha_2 + 9\alpha_1^3 - 320\alpha_2^2\omega^2 - 768\alpha_1\alpha_2\omega^2 - 64\alpha_1^2\omega^2 + 384\alpha_2\omega^4 + 128\alpha_1\omega^4\right.\right.$$

$$+ 4\sqrt{\alpha_1\alpha_2}\left(81\alpha_2^2 + 162\alpha_1\alpha_2 + 27\alpha_1^2 - 256\alpha_2\omega^2 - 128\alpha_1\omega^2 + 128\omega^4\right)\cos\left(\left(\beta_1 + \beta_2\right)\omega\right) + 6\alpha_1\alpha_2\left(45\alpha_2 + 27\alpha_1 - 64\omega^2\right)$$

$$\left.\left. \times \cos\left(2\left(\beta_1 + \beta_2\right)\omega\right) + 72\alpha_1\alpha_2\sqrt{\alpha_1\alpha_2}\cos\left(3\left(\beta_1 + \beta_2\right)\omega\right)\right\}\sin\left(\left(\beta_1 + \beta_2\right)\omega\right)\right]$$

$$+ \varepsilon^4\left[\frac{1}{786432\sqrt{\alpha_2}\omega^{12}}\left\{-4\sqrt{\alpha_1}\left(2527\alpha_1^4 + 3\alpha_1^3\left(12635\alpha_2 - 5032\omega^2\right) + \alpha_1^2\left(-90576\alpha_2\omega^2 + 24416\omega^4\right) + \alpha_1\left(-88445\alpha_2^2\right.\right.\right.\right.$$

$$\left.+ 150960\alpha_2^2\omega^2 - 8448\omega^6\right) + \alpha_2\left(-22743\alpha_2^3 + 105672\alpha_2^2\omega^2 - 122080\alpha_2\omega^4 + 25344\omega^6\right)\right)\cos\left(\left(\beta_1 + \beta_2\right)\omega\right)$$

$$+ \sqrt{\alpha_2}\left(-37905\alpha_1^4 - 101080\alpha_1^3\alpha_2 + 151620\alpha_1^2\alpha_2^2 + 151620\alpha_1\alpha_2^3 + 12635\alpha_2^4 + 161024\alpha_1^3\omega^2 - 483072\alpha_1\alpha_2\omega^2 - 80512\alpha_2^3\omega^2\right.$$

$$- 146496\alpha_1^2\omega^4 + 292992\alpha_1\alpha_2\omega^4 + 146496\alpha_2^2\omega^4 - 67584\alpha_2\omega^6 + 6144\omega^8 - 4\alpha_1\left(7581\alpha_1^4 + \alpha_1^2\left(25270\alpha_2 - 30192\omega^2\right)\right.$$

$$\left.-7\alpha_1\left(5415\alpha_2^2 - 3488\omega^4\right) - 4\alpha_2\left(7581\alpha_2^2 - 22644\alpha_2\omega^2 + 12208\omega^4\right)\right)\cos\left(2\left(\beta_1 + \beta_2\right)\omega\right) - 4\alpha_1\sqrt{\alpha_1\alpha_2}\left(7581\alpha_1^2 - 17689\alpha_2^2\right.$$

$$\left.\left.\left.\left. -15096\alpha_1\omega^2 + 25160\alpha_2\omega^2\right)\cos\left(3\left(\beta_1 + \beta_2\right)\omega\right) + \left(-10108\alpha_1^3\alpha_2 + 15162\alpha_1^2\alpha_2^2\right)\cos\left(4\left(\beta_1 + \beta_2\right)\omega\right)\right)\right\}\right].$$

$$\text{(B4)}$$